# GROUND-BASED CISLUNAR SPACE SURVEILLANCE DEMONSTRATIONS AT LOS ALAMOS NATIONAL LABORATORY

Y. Sechrest[*], M. Vance[*], C. Ward[*], W. Priedhorsky[*], R. Hill[*], M. Giblin[*], W. T. Vestrand[*], P. Wozniak[*]

Surveillance of objects in the cislunar domain is challenging due primarily to the large distances (10x the Geosynchronous orbit radius) and total volume of space to be covered. Ground-based electro-optical observations are further hindered by high background levels due to scattered moonlight. In this paper, we report on ground-based demonstrations of space surveillance for targets in the cislunar domain exploiting the remarkable performance of 36cm, F2.2 class telescopes equipped with current generation, back-side illuminated, full-frame CMOS imager. The demonstrations leverage advantageous viewing conditions for the Artemis Orion vehicle during its return to earth, and the total lunar eclipse of November 8th 2022 for viewing the CAPSTONE vehicle. Estimated g-band magnitudes for vehicles were 19.57 at a range of 4.4e5 km and 15.53 at a range of 3.2e5 km for CAPSTONE and Artemis Orion, respectively. In addition to observations, we present reflectance signature modeling implemented in the LunaTK space-sensing simulation framework and compare calculated apparent magnitudes to observed. Design of RApid Telescopes for Optical Response (RAPTOR) instruments, observing campaigns of Artemis Orion and CAPSTONE missions, and initial comparisons to electro-optical modeling are reviewed.

**INTRODUCTION**

With at least four space-faring nations and one commercial entity having announced missions to cislunar space in the next decade and plans for renewed human presence on the moon, there exists a steadily growing need for surveillance of this domain to secure assets and personnel.[†,1] Recognizing this need, we have undertaken a comprehensive study of Electro-Optical (EO) space sensing in the cislunar domain targeted at furthering understanding of sensing constraints and performance. As part of this initiative, a new generation of Rapid Telescopes for Optical Response (RAPTOR) recently participated in observing campaigns to detect the Artemis-Orion and CAPSTONE missions at ranges comparable to the average Earth-Moon distance.[2] Three instruments participated in observations: two located at Fenton Hill Observatory (FHO) and one located at the Los Alamos Neutron Science Center. Geographic separation of the two sites is sufficient to support optical ranging at distances comparable to the average Earth-Moon distance. In addition to observing campaigns, our team is developing the Luna ToolKit (LunaTK) simulation framework for simulating multi-agent space sensing scenarios in the cislunar domain with coordinated tip-and-cue response behaviors. Included in this framework is a multi-fidelity EO simulation tool which implements both simple, spherical reflector models and more complicated facet-based models for simulating EO signatures of objects. While initial studies have focused on EO sensing, LunaTK is intended to be extensible to other sensor modalities. In this paper, we review the results of the observing campaigns and compare observations with optical simulations.

---

[*] Los Alamos National Laboratory, P.O. Box 1663 Los Alamos, NM 87545
[†] https://www.bbc.com/news/science-environment-64002977

**RAPID TELESCOPES FOR OPTICAL RESPONSE (RAPTOR)**

The original RAPTOR network operated in the early 2000's and consisted of a pair of robotic telescopes (RAPTOR-A and RAPTOR-B) constructed at two geographically separated sites: one at Fenton Hill Observatory (FHO) in the Jemez Mountains and the other at the Los Alamos Neutron Science Center (LANSCE) just outside of Los Alamos townsite.[2] These instruments consisted of 4 CCD imagers with 85mm / F1.8 lenses flanking a central CCD imager with a 400mm / F2.8 lens. The instruments were originally designed to monitor the night sky of the northern hemisphere in real time searching for astronomical transients and proved to be quite successful for observing prompt optical emission and optical afterglow of gamma ray burst (GRB) events.[3] Through the 2000's and 2010's the RAPTOR network made several high impact observations of GRB events and provided significant insight into the physical processes of these high-energy astronomical events.

**Table 1 RAPTOR-R Instrument Design Metrics**

| | | | | | |
|---|---|---|---|---|---|
| Focal Length [mm] | 790 | Width [deg.] | 2.6 | [pix] | 9600 |
| Aperture Dia. [mm] | 360 | Height [deg.] | 1.7 | [pix] | 6422 |
| F-Number | 2.2 | Area [sq. deg.] | 4.5 | [Mpix] | 61 |
| Slew Rate [deg./sec] | 50 | Platescale ["/pix] | 0.98 | ["/mm] | 261 |
| Track Rate [deg./sec] | 5 | Sensor Diagonal [mm] | 43 | | |

The new generation of RAPTOR instruments are designed to leverage recent advancements in commercially available technologies: back-side illuminated full-frame CMOS detectors, advanced wide-field Rowe-Ackermann Schmidt Astrographs (RASA), and fast slewing direct drive mounts. These 360mm / F2.2 instruments were commissioned for studies of transient astronomical events, including gravitational wave follow-up and sub-second optical flashes, and for other space domain awareness (SDA) applications. The instrument design is summarized in Table 1. Three instruments have been constructed: one located at the LANSCE site (RAPTOR-B), and two located FHO (RAPTOR-A, RAPTOR-R). The LANSCE and FHO sites are separated by approximately 38km East-West and provide sufficient parallax to optically range targets at distances beyond Geosynchronous Earth Orbit (GEO) radii. Astrophotography grade, cooled CMOS cameras by manufacturers QHY and ZWO, based on the SONY IMX455 sensor, are used for their high peak quantum efficiency (90% at 500nm), full-frame 43mm sensor diagonal, and small 3.76 micron pixel size. Paired with the 360mm aperture RASA, this yields a 4.5 square degree field of view with 0.98 arcsec per pixel resolution. Figure 1(left) shows an example image of galaxy M31 taken from RAPTOR-R illustrating image dimensions.

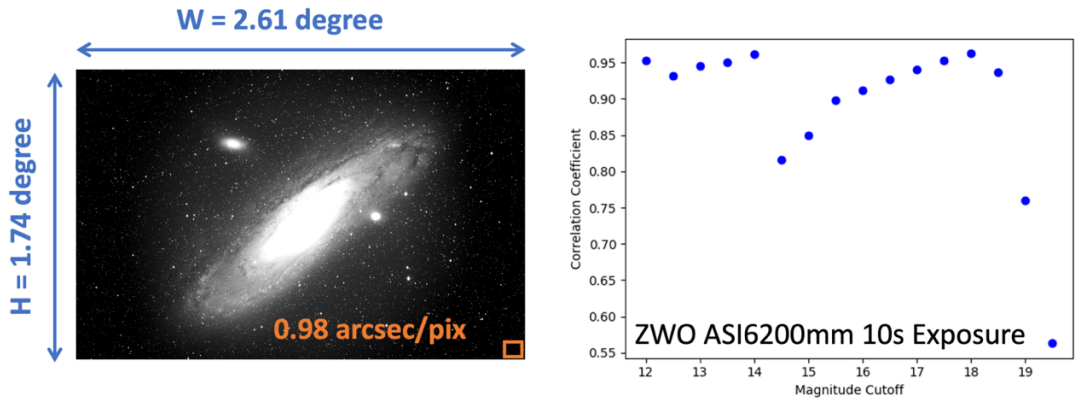

**Figure 1. (left) Example field of view for new RAPTOR instruments. (right) Correlation of instrumental to catalog magnitude as a function of magnitude cutoff for a 10 second exposure.**

With dark sky conditions at FHO, the unfiltered monochrome instruments achieve a limiting magnitude of approximately 18th magnitude (R-band) in a 10 second exposure. Limiting magnitude was estimated by correlation of instrumental magnitudes of detected sources with USNO-A2 R-band magnitudes for sources in the catalog that match observed RA/DEC positions. Figure 1 plots the correlation with catalog as a function of magnitude cutoff showing a steep decline above 18th magnitude (R-Band).

**OBSERVATIONS OF CAPSTONE AND ARTEMIS-ORION**

The Artemis Orion vehicle (approximate dimensions of crew and service module: 16.5ft diameter x 26.7ft length) was observed on December 8th 2022 during the return to earth at a range of 3.2e5 km, and the CAPSTONE vehicle (12U CubeSat) was observed during the total lunar eclipse on November 8th 2022 at a range of 4.4e5 km.[*] Both vehicles were observed with the RAPTOR-R instrument at FHO, and Artemis Orion was additionally observed stereoscopically from FHO and LANSCE using instruments RAPTOR-A and RAPTOR-B. This allowed our team to estimate the range of the capsule optically using the parallax between the two sites, and these results have been reported on previously.[4] Details of the observations for both vehicles are presented below.

---

[*] https://ssd.jpl.nasa.gov/horizons/

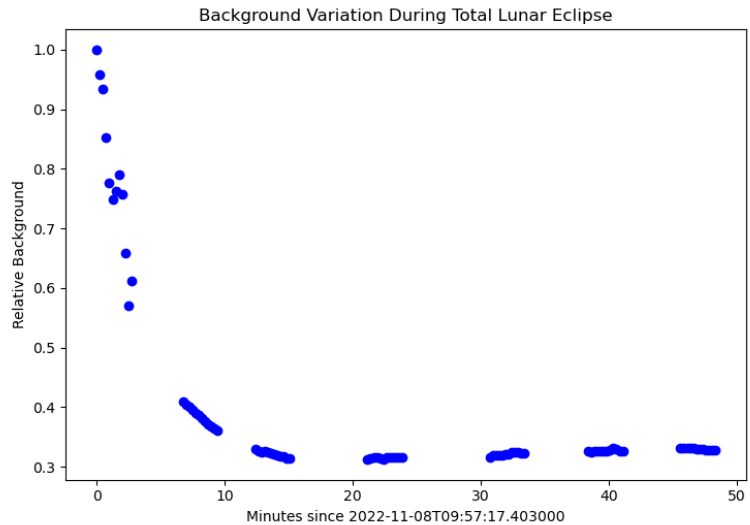

**Figure 2 Measured background signal relative start of observation time: 2022-11-08T09:57+00:00 near pointing of RA 49.6 º Dec 5.7 º as observed from FHO. Total eclipse begins at t=19 minutes.**

The total lunar eclipse of November 8th 2022 presented a unique opportunity for ground-based observation of targets in the cislunar environment. Time windows for various stages of the eclipse were: penumbral (8:02-13:56+00:00), partial (9:09-12:49+00:00), total (10:16-11:41+00:00).[*] The nature of the eclipse geometry ensured near optimal observing conditions that are not normally realizable under other conditions; the Sun-Vehicle-Observer (SVO) phase angle for targets near the moon was low which ensures near maximal diffuse reflection signature, and the sky background is much reduced due to the reduction of scattered moonlight. Figure 2 shows measured background counts near pointing of RA 49.6º Dec 5.7º during the observation window (2022-11-08T10:00+00:00 to 2022-11-08T11:00+00:00). The plot spans from 20 min prior to total eclipse until 30 min after start of total eclipse and captures the change in background during transition from partial to total eclipse. Note, however, that background levels even at t=0 minutes are significantly reduced from background levels prior to and after eclipse. Beyond 11:00+00:00, viewing of the moon from FHO was obstructed by intermittent cloud cover.

The CAPSTONE vehicle's motion during sidereal rate tracking was estimated at about 2.5 pixels (about one PSF width) in a 10 second exposure, but the apparent magnitude was estimated to be below the single frame magnitude limit during the eclipse. To improve the detectability of the CAPSTONE vehicle, sequences of twelve 10 second exposures were taken while tracking at sidereal rate near the location of the predicted ephemeris. These were then combined using a simplified drizzle-like image stacking algorithm that resembles a shift and add approach.[5] Frame-to-frame shifts were determined by the CAPSTONE ephemeris reported by the JPL Horizons database.[†] Image stacking reveals an object within 40" of the CAPSTONE astrometric ephemeris location with similar motion to the provided ephemeris during the observation period. The standard deviation of the background is reduced by image stacking such that the average peak counts of the object, which is about 1 standard deviation above background for a single frame, exceeds 5 standard deviations when 12 images are combined (Figure 3). A faint point source like object was detected in four independent 12 frame stacks taken between 2022-11-08T10:00+00:00 and 2022-11-08T11:00+00:00 with the same angular offset from the CAPSTONE ephemeris in each stack. The object was also detected in permutations of stacks with a single frame removed which indicates it is not due to sporadic high counts at the location for a single frame (*e.g.* a cosmic ray).

---

[*] https://earthsky.org/astronomy-essentials/total-lunar-eclipse-nov8-2022/

[†] https://ssd.jpl.nasa.gov/horizons/

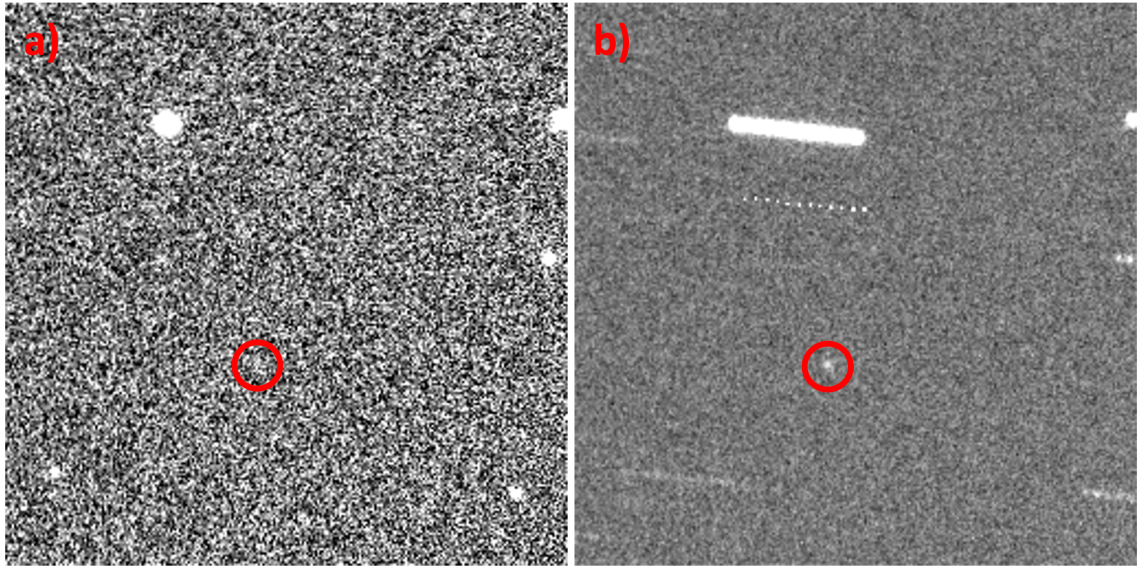

**Figure 3 Single frame (a) and 12 frame stack (b) of 10 second exposures.**

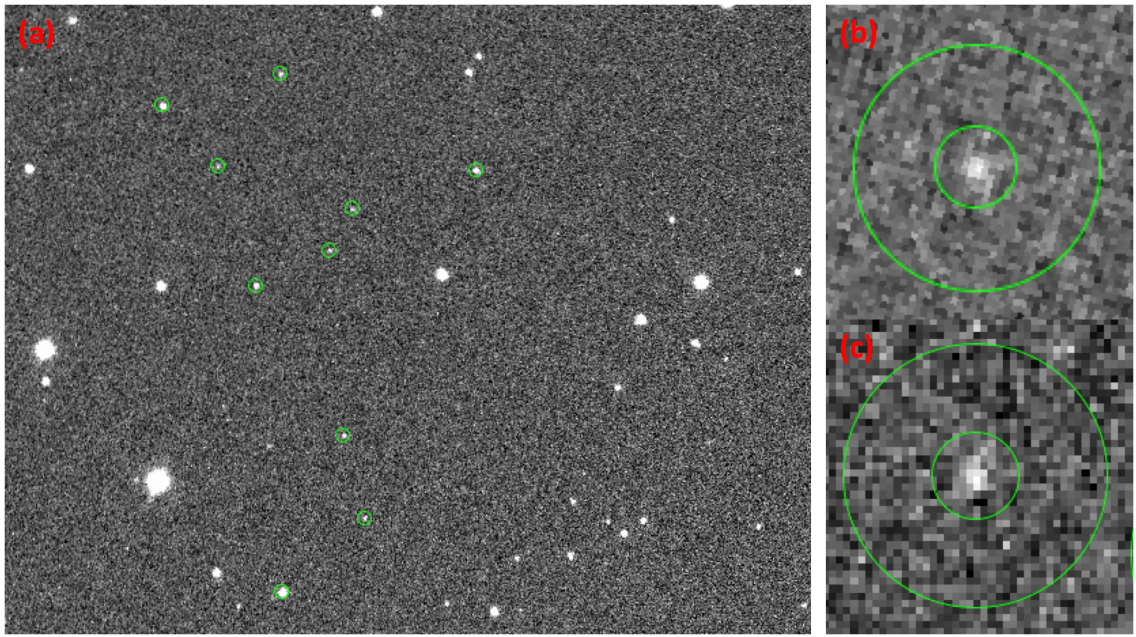

**Figure 4 (a) Reference single frame exposure near RA 49.6 ° Dec 5.7 ° with reference stars indicated by green circles. Reference apertures for (b) CAPSTONE 12-frame stack and (c) reference star at RA 49.66 ° Dec 5.75 °**

To estimate the object's apparent magnitude, we apply aperture photometry to a set of nearby reference stars within one 10 second exposure and compare with Sloan Digital Sky Survey (SDSS) photometry.[6] The reference image with selected reference stars highlighted is shown in Figure 4 beside apertures for (b) CAPSTONE and (c) a reference star at location RA 49.66 ° Dec 5.75 °. Reference star locations, measured signal and SDSS u, g, and r magnitudes are given in Table 2 with measured signal of CAPSTONE from a 12 frame stack. The images taken by RAPTOR-R are unfiltered, monochrome images, so to compare with photometry of reference sources, we first estimate a fit to the instrumental magnitude of the form:

$$I = Ag + B(g-r) + C(g-u) + D \qquad (1)$$

with fit parameters *A*=0.986, *B*=-0.898, *C*=0.046, *D*=-14.449. Rearranging the above equation to solve for g, plugging in the measured instrumental magnitude for CAPSTONE, and assuming the reflected light follows solar emission ($g - r = 0.44, g - u = -1.43$), we obtain the apparent magnitude of the object referenced to Sloan g-band to be 19.57+/- 0.02 magnitudes (uncertainty is estimated from fit to reference stars alone). This is ~0.5 magnitudes fainter than the reference stars in the image that were near the single frame detection limit for a 10 second exposure.

Table 2 Table of reference star locations, integrated counts, and Sloan Digital Sky Survey photometry for u, g, and r bands. Measured counts for CAPSTONE stack are also included.

| RA (deg.) | DEC (deg.) | Signal (counts) | Sloan u (mag.) | Sloan g (mag.) | Sloan r (mag.) |
|---|---|---|---|---|---|
| 49.6841 | 5.7317 | 1888.33 | 18.63 | 17.36 | 16.79 |
| 49.6662 | 5.7400 | 866.08 | 21.19 | 18.7 | 17.64 |
| 49.6605 | 5.7499 | 688.76 | 21.18 | 19.01 | 17.99 |
| 49.6306 | 5.7590 | 2363.2 | 18.49 | 17.2 | 16.65 |
| 49.6778 | 5.7826 | 1089.7 | 21.71 | 19.02 | 17.42 |
| 49.6931 | 5.7605 | 653.42 | 20.02 | 18.51 | 17.84 |
| 49.7064 | 5.7752 | 2745.18 | 20.43 | 17.7 | 16.3 |
| 49.6632 | 5.6954 | 1042.18 | 20.67 | 18.51 | 17.58 |
| 49.6584 | 5.6754 | 973.94 | 19.96 | 18.24 | 17.49 |
| 49.6784 | 5.6577 | 13274.44 | 17.39 | 15.53 | 14.67 |
| CAPSTONE | | | | | |
| 49.6494 | 5.7074 | 231.64 | | | |

Being much larger than CAPSTONE, the Artemis Orion vehicle is a less challenging target for the RAPTOR telescopes, and the vehicle is easily detected in a single 10 second exposure while tracking at sidereal rate. Figure 5 shows two 10 second exposures taken 32 seconds apart with the starting and ending positions of the Artemis Orion vehicle marked by red and green arrows, respectively. Estimated angular velocity at time of observation is 16" per minute. A similar fitting procedure was carried out to estimate the apparent magnitude of this vehicle and the following parameters were determined: *A*=0.971, *B*=-0.638, *C*=-0.433, *D*=-12.668. Using these parameters and the color terms for the Sun, the g-band apparent magnitude was estimated to be 15.53 +/- 0.16 magnitudes.

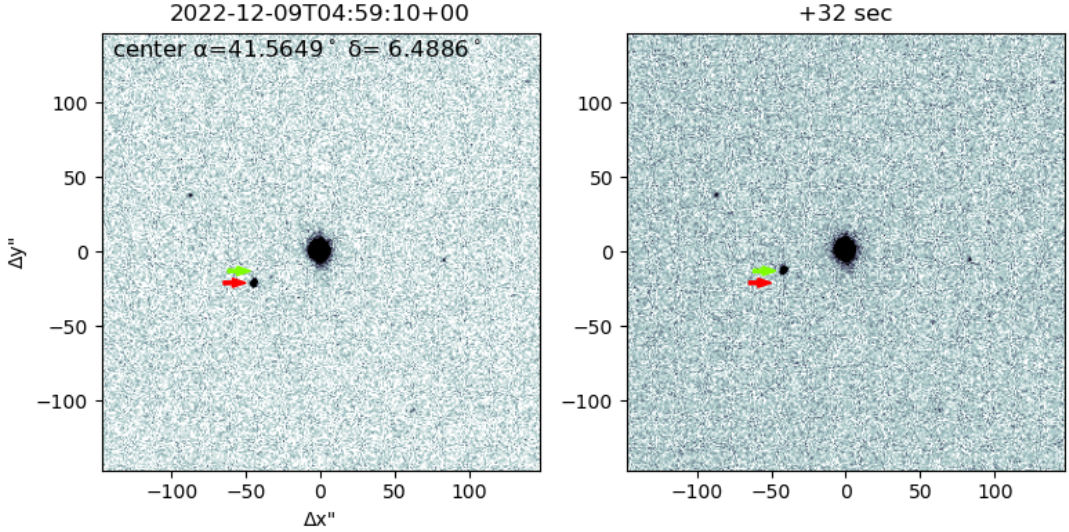

*Figure 5 Two 10 second exposures taken 32 seconds apart during observing of Artemis-Orion capsule. Red and Green arrows show vehicle position for first and second image in sequence respectively.*

**LUNATK ELECTRO-OPTICAL SENSING MODELS**

In addition to instrumentation and observation work at FHO and LANSCE, our team is also engaged in developing a comprehensive simulation framework for sensing in the cislunar domain named Luna ToolKit (LunaTK).[7] The framework seeks to integrate astrodynamics libraries, varying fidelity sensor models, and dynamic agent-based behaviors to simulate sensing scenarios in the cislunar domain. As part of the greater simulation framework, we have implemented both simple spherical reflector models and basic facet models to represent passive EO sources, and we are developing a range of sensor models including pixelated CMOS, event-based, and photon counting sensors. While initial development has primarily focused on passive EO sensing, the tools are intended to be extensible to other sensing modalities as well.

The EO sensing models employed in LunaTK begin with an illumination of a target object by an incoming irradiance (W/m$^2$). The target to illumination and target to sensor distances are assumed to be much greater than the target dimensions so that incoming and outgoing rays are parallel for all points on the body. A target may be represented by a spherical reflector with mixed diffuse and specular components, or by a facet model with facets assigned empirical or physically plausible Bidirectional Reflectance Distribution Functions (BRDFs). The contributions to outgoing radiant intensity (W/sr) in the direction of the sensor is then integrated for all facets(points) on the body(sphere) according to the rendering equation.[8] For a spherical reflector, the outgoing radiant intensity is given by:

$$\frac{d\Phi_e}{d\omega} = \frac{2a_d r^2}{3\pi} I_\odot (\sin\phi - (\phi - \pi)\cos\phi) + \frac{a_s r^2}{4} I_\odot \qquad (2)$$

where $a_d$ and $a_s$ are the diffuse and specular albedo, $r$ is the radius of the sphere, $I_\odot$ is the solar irradiance, and $\phi$ is the phase angle subtended by the sun and observer from the point of the target. For a faceted body the outgoing radiant intensity in direction $\omega_o$ due to light incident from $\omega_i$ is given by:

$$\frac{d\Phi_e}{d\omega} = \sum_j g(\vec{x}^j, \omega_o, \omega_i) f(\vec{x}^j, \omega_o, \omega_i) I_\odot (\omega_i \cdot \hat{n}^j)(\omega_o \cdot \hat{n}^j) A_{facet}^j \qquad (3)$$

where $g$ is the geometric shadowing factor for the facet, $f$ is the BRDF, $\hat{n}$ is the facet normal, $A$ is the area of the facet, and $x$ is the position on the body of the facet. This is then propagated to the sensor location and multiplied by the projected area of the collecting aperture to calculate the received radiant flux (W or photons/s) at the sensor. Additional simplifying assumptions are currently made in the development of sensing models: the body is assumed to not change shape. Secondary reflections are not modeled but simple shadowing and obscuration is modeled. Material assignments are 'best guess', and BRDFs are isotropic, effectively single-wavelength, without treatment of polarization. Results presented in this section use Maxwell-Beard BRDF models from the SatAC Materials Database (Version 5.6, February 08, 2011).[9]

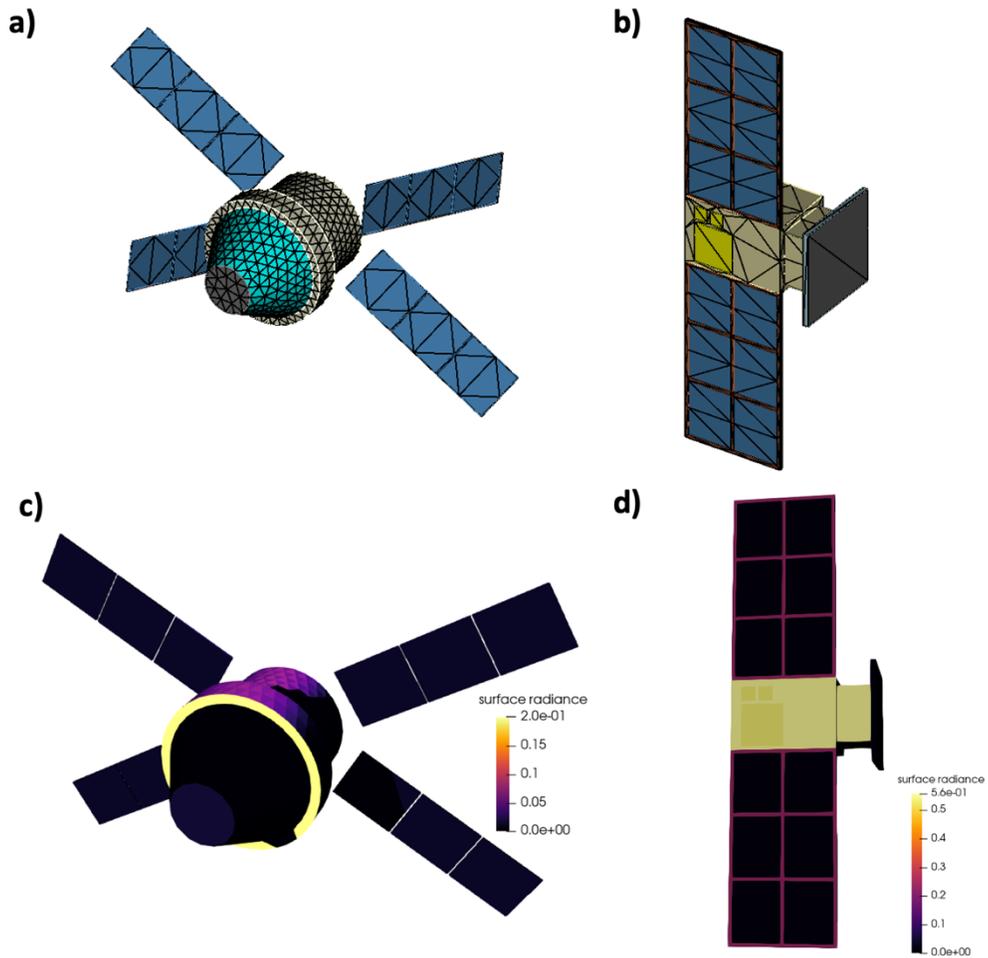

**Figure 6 Facet models for simplified Artemis (a) and CAPSTONE (b) and models shaded by facet contribution to reflected radiance normalized by facet area for Artemis (c) and CAPSTONE (d) example viewing geometries.**

The observations of CAPSTONE and the Artemis Orion vehicles provide important data that can be used to validate LunaTK sensing models against real world scenarios. Simple facet models for Artemis Orion and CAPSTONE have been built from publicly available data about these missions, and these are shown in Figure 6 (a) and (b) respectively. Figure 6 (c) and (d) show facet models shaded by facet contribution to reflected radiance (normalized by area of facet) for one example viewing geometry (*i.e.* pair of incoming and outgoing rays). For comparison with estimated g-band magnitudes, the integrated radiant intensity is propagated a distance R from satellite to top of atmosphere and converted to magnitude:

$$g = -26.47 - 2.5 \, LOG_{10} \left( \frac{d\Phi_e}{d\omega} \cdot \frac{1}{I_\odot R^2} \right) \tag{4}$$

For the SDSS g filter band, the apparent magnitude of the sun is taken to be -26.47 and the solar irradiance is $I_\odot \approx 200 \, W/m^2$.[10,11,12]

Calculations of g-band magnitudes for Artemis Orion and CAPSTONE vehicles are presented and Figure 7 and Figure 8. Figure 7 plots the CAPSTONE and Artemis calculated magnitudes as a function of SVO phase angle for spherical reflector and facet model. The observational data is included on the plots at the phase angle corresponding to time of observation estimated from ephemeris data (CAPSTONE=12.5°, Artemis=36.5°). The satellite attitude at time of observation is not known, so an attitude is chosen that best corresponds to the observational data. Viewing geometry corresponds to those shown in Figure 6 (c) and (d). For CAPSTONE, the vehicle is oriented with solar panels directed toward the sun and incoming and outgoing vectors rotated about the model polar axis by an arbitrary roll angle. With nose of crew capsule cone defining the model polar axis, the Artemis model polar angle was rotated 28 degrees so that nose pointed below plane defined by incoming and outgoing rays. Calculated g-band magnitudes for these orientations were 19.83 for CAPSTONE and 15.53 for Artemis, and equivalent Lambertian sphere diameters were 0.67m and 3.45m, respectively.

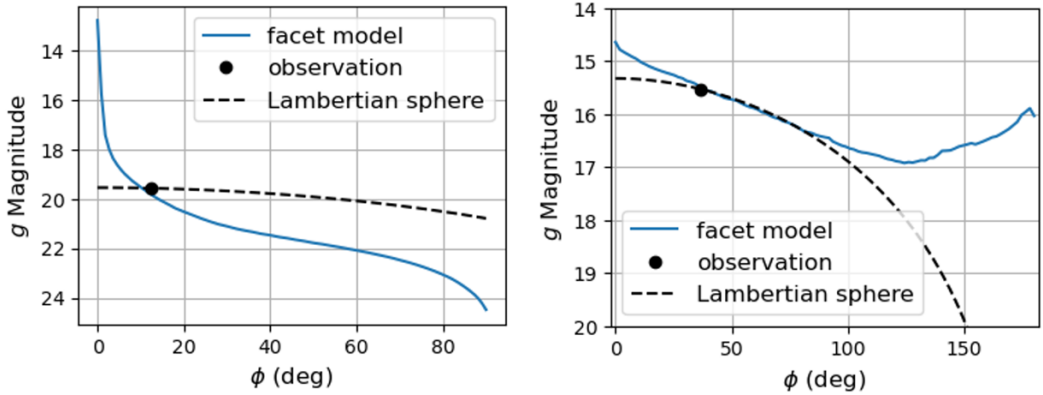

**Figure 7 Plots of estimated g-band magnitude versus SVO phase angle for CAPSTONE (left) and Artemis Orion (right) with selected attitude and material BRDFs.**

To capture the variability in estimates of g-band magnitudes, many calculations are performed for the CAPSTONE model while varying orientation of vehicle at fixed SVO angle. The half-angle direction vector (bisecting incoming and outgoing rays) is rotated by polar (Φ) and azimuthal (Θ) angles with a fixed, arbitrary roll about the half-angle vector. Results are presented in Figure 8. The half-angle direction vectors are sampled on a unit sphere according to a HEALPix grid with approximately one degree resolution (nside = 128).[13] The CAPSTONE vehicle predominantly has cubic symmetry, so the outgoing radiance exhibits specular lobes around each of the primary faces of the model except the face at $\Phi = 90^o, \Theta = 0^o$ which is oriented about the S-band antenna. Outgoing radiance from this face is suppressed due to choice of material BRDF applied to this facet.

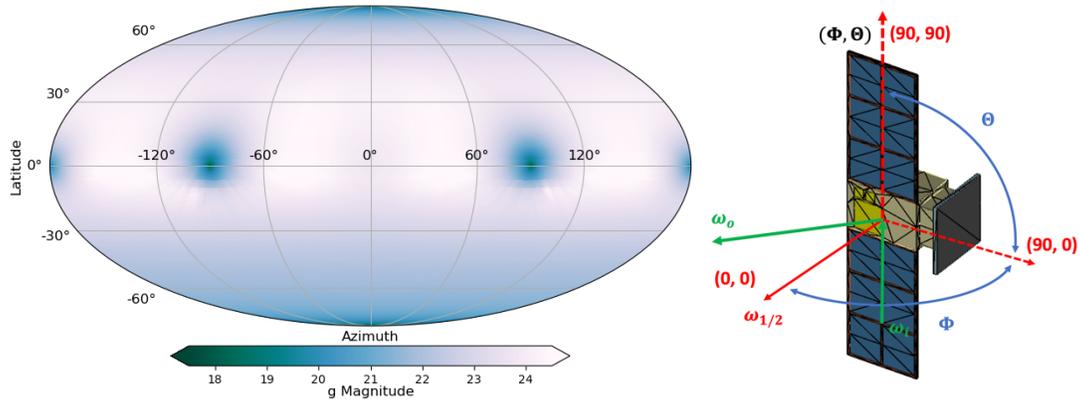

**Figure 8** Contour plot (left) of calculated g-band magnitudes for CAPSTONE facet model. For fixed SVO phase angle of 12.5º and fixed roll of incoming and outgoing rays about spacecraft polar axis, polar angle ($\Phi$) and azimuth ($\Theta$) of half-angle vector is scanned as shown in figure at right. Latitude is defined as $90^o - \Phi$.

## CONCLUSION

In Fall and Winter of 2022, the RAPTOR instrument network at LANL was used to perform a series of ground-based, electro-optical sensing demonstrations of targets in cislunar space. The RAPTOR network consists of three 36cm aperture, F2.2 class optical telescopes with one instrument located on the Los Alamos Neutron Science Center mesa near Los Alamos, NM and two instruments located at Fenton Hill Observatory west of Los Alamos in the Jemez Mountains. The instruments were designed to leverage low-cost, commercial off-the-shelf technologies for imager (cooled, back-illuminated CMOS sensors), optical tube assembly (Rowe-Ackermann-Schmidt Astrograph), and mount (fast slewing, direct drive). Using these instruments, our team observed the CAPSTONE vehicle during the total lunar eclipse on November 8th 2022, and the Artemis Orion vehicle during its return to earth on December 8th 2022. Estimated g-band magnitudes for the objects were 19.57 +/- 0.02 at a range of 4.4e5 km and 15.53 +/- 0.16 at a range of 3.2e5 km for CAPSTONE and Artemis Orion, respectively.[*] Calculations of g-band magnitude were performed under various viewing geometries using LunaTK facet models, and observational results were found to be in good agreement given a plausible choice for unknown vehicle attitude at the time of observation (*e.g.* solar panels directed toward sun). More careful analysis is warranted but will require accurately specifying vehicle attitude at time of observation.

## ACKNOWLEDGMENTS

This work was support by Los Alamos National Laboratory Directed Research and Development funds, project 20220798DI.

---

[*] https://ssd.jpl.nasa.gov/horizons/